\documentstyle[aps,prl,epsf]{revtex}

\def\la{{\lambda}}
\def\lae{{\la_{\rm eff}}}
\def\be{{\beta             }}
\def\De{{\Delta            }}
\def\hh{{\hat h            }}
\def\hs{{\hat s            }}
\def\hd{{\hat d            }}
\def\ha{{\mbox{${1\over2}$}}}
\def\nl{{\newline          }}
\def\noi{{\noindent        }}
\def\lgl{\left\langle                }
\def\rgl{\right\rangle               }
\def\dll{\left\langle \!\!\left\langle }
\def\drl{\right\rangle\!\!\right\rangle}
\def\dle{\left\langle \! \left\langle }
\def\dre{\right\rangle\! \right\rangle}
\def\beq{\begin{equation} }
\def\eeq{\end{equation}   }

\begin{document}

\title{Swollen-Collapsed Transition in Random Hetero-polymers}

\author{ A.~Trovato, J.~van~Mourik and A.~Maritan }

\vskip 1cm

\address{International School for Advanced Studies (SISSA),\\
and Istituto Nazionale di Fisica della Materia,\\
Via Beirut 2-4, 34014 Trieste, Italy }

\address{Abdus Salam International Center for Theoretical Physics,\\
Strada Costiera 11, 34100 Trieste (Italy) }

\maketitle

\begin{abstract}

\section*{Abstract}

A lattice model of a hetero-polymer with random hydrophilic-hydrophobic
charges interacting with the solvent is introduced, whose continuum
counterpart has been proposed by T.~Garel, L.~Leibler and H.~Orland
\cite{GLO}.
The transfer matrix technique is
used to study various constrained annealed systems which approximate at
various degrees of accuracy the original quenched model. For highly
hydrophobic chains an ordinary
$\theta$-point transition is found from a high temperature swollen phase
to a low temperature compact phase. Depending on the type of constrained
averages, at very low temperatures a swollen phase or a coexistence between
compact and swollen phases are found. The results are carefully compared
with the corresponding ones obtained in the continuum limit, and various
improvements in the original calculations are discussed.

PACS numbers: 05.70.fh 61.41.+e 64.75.+g 75.10.nr

\end{abstract}


\section{Introduction}

The main reason for the study of random hetero-polymers in solutions, is a
possible connection of this problem with the protein folding problem
\cite{BryWol,GarOrl,ShaGut}. Indeed, proteins are believed to be by nature
selected special
cases of random hetero-polymers. Before dealing with these special cases,
it is of great importance to understand the typical behaviour of the
various kinds of random hetero-polymer models that have been introduced, as
it may give important insight in which types of interactions are
indispensible for folding, and which types of interactions, on the other
hand, are of secondary importance.

Several models of (quenched) randomness have been considered. Here, we
study the role of the solvent (water) in the equilibrium \cite{Anf}
properties of the collapsed phase, as it is
commonly believed that the hydrophobic effect \cite{Dill} is the
main driving force for the folding transition. Most proteins in nature
consist of a strongly hydrophobic core, surrounded by hydrophilic (less
hydrophobic) residues. We restrict ourselves to the simple coarse grained
model, that was originally introduced by Obukhov \cite{Ob}, where the
monomers of a single chain are randomly hydrophilic or hydrophobic (RHH), and
interact with the solvent molecules through an effective two-body short
range interaction.
The statics of the continuum version of this model has been studied by
Garel {\it et al.} \cite{GLO}, both in the case of annealed and quenched
disorder, while the dynamics (with quenched disorder) has been studied by
Thirumalai {\it et al.} \cite{TAB}. The model has been studied also
by Timoshenko {\it et al.} \cite{Tim} and Moskalenko {\it et al.} \cite{Mos}
with the Gaussian self-consistent method.

We have choosen to study a (2d square) lattice version of the RHH model,
and the method we used to assess the conformational entropy, is the
transfer matrix (TM) method, which is most fit to study the case
of annealed disorder. Furthermore, using the approximation scheme
introduced by Morita \cite{Mo}, we are able to give lower bounds for the
quenched free energy. It will turn out that the annealed case may exhibit a
very rich phase diagram and re-entrant compact-swollen transitions, and we
come to different conclusion than Garel {\it et al.} \cite{GLO}. The case
of the annealed average with fixed mean for the hydrophobic-hydrophilic
charges, gives the same results one can get in the continuum limit for the
quenched average using the method of reference \cite{GLO}
(see section \ref{disc} for details). We go one step forward analyzing a
better approximation to the quenched system which cures some problems present
in the previous approximations and in the standard approach presented in
\cite{GLO}.

This work is built up in the following way.
In section \ref{defmod}, we introduce the model in more detail.
In section \ref{ConAnn}, we introduce the concept of constrained annealing; in
section \ref{phdg} we show that the effective models after averaging over the
disorder involve $2-$ and $3$-body interactions and the general
phase diagram of such kind of models is discussed.
In section \ref{tm}, the TM method is used to assess the
conformational entropy of the polymer chain. The results are presented in
section \ref{res}, together with an outlook of the items that are still to be
investigated.
Finally, in section \ref{disc} we give an interpretation of our results,
and a detailed comparison with those obtained for the continuum model by
\cite{GLO}.

\section{Definition of the Model}
\label{defmod}

The polymer chain is represented by a self-avoiding walk (SAW) on a lattice
where each site is either visited by the walk (i.e. is occupied
by one monomer of the chain), or occupied by a solvent molecule.
The interactions in the model are two-body short-range interactions. The
only interactions we take into account, are those between solvent
molecules and monomers if they occupy nearest-neighbor sites.
Hydrophilicities $\la_i$ are attached to each monomer $i$ of the walk
such that the Hamiltonian is given by

\beq
{\cal H}=-\sum_{i=0}^N\la_i z_i\ ,
\label{H}
\eeq

\noi where the sum runs over the $N+1$ sites of the lattice occupied
by the $N$-step walk, and $z_i$ is the number of nearest-neighbor
contacts of monomer $i$ with solvent molecules, i.e. the number of
nearest-neighbor sites of site $i$ not occupied by the walk.

The hydrophilicities $\la_i$ are supposed to be independent
identically distributed random variables with a Gaussian distribution with
mean $\la_0$ and variance $\la$:

\beq
P(\la_i)={1\over\sqrt{2\pi\la^2}}
\exp\left(-{(\la_i-\la_0)^2\over2\la^2}\right)\ ,
\label{P}
\eeq

\noi and the average over this (a priori) distribution is indicated by
$\dle\cdot\dre$ . If $\la_i>0$, the corresponding monomer is
hydrophilic and attracts solvent molecules, whereas if $\la_i<0$, the
monomer is hydrophobic and repels solvent molecules.

The canonical partition function for SAW of $N$ steps with a fixed
disorder configuration $\{\la_i\}$ is then

\beq
{\cal Z}_N(\{\la_i\})=\sum_{W_N}
\exp\left(\be\sum_{i=0}^N\la_i z_i\right)\ ,
\label{Z}
\eeq

\noi where the sum has to be taken over all $N$-step SAW starting from
the origin.

If monomers can rearrange themselves along the chain and change their
hydrophilicities, e.g. with chemical reactions, these have to be considered
as thermal annealed variables, which approach equilibrium in the same time
scale as the configurational degrees of freedom. The physics of such
hetero-polymers is given by the average of the partition function over the
disorder distribution (annealed average)

\beq
\dle{\cal Z}_N(\{\la_i\})\dre=\sum_{W_N }\exp\left(\be\la_0
\sum_{i=0}^Nz_i+{\be^2\la^2\over2}\sum_{i=0}^Nz_i^2\right)\ .
\label{mZ}
\eeq

\noi
Instead, if the monomer sequence of the chain is fixed, as it is the case for
proteins, the hydrophilicities are frozen while the polymer is approaching
thermal equilibrium; the average over the disorder distribution has then to
be taken over the logarithm of the partition sum (quenched average)
\cite{Bro}, to yield the quenched free energy

\beq
f_q=-{1\over\be}\dle\ln[{\cal Z}_N(\{\la_i\})]\dre\ ,
\label{fq}
\eeq

\noi which is a much harder task to accomplish.

\section{Constrained Annealing}
\label{ConAnn}

In order to avoid the difficult direct computation of the quenched average
(\ref{fq}), we have applied an idea first introduced by Morita
\cite{Mo}. This is the so-called Equilibrium Ensemble Approach (EEA) (see
e.g. \cite{Ku} for a recent review and discussion). The EEA consists of a
systematic approximation procedure for the quenched free energy by annealed
averages. It can be shown \cite{Ku} that each successive approximation gives
a better or equally good lower bound for the quenched free energy.

Each approximation consists in performing an annealed average over a new
Hamiltonian ${\cal H}^*\equiv{\cal H}+{\cal H}_d$, where ${\cal H}$ is the
original Hamiltonian (\ref{H}), and ${\cal H}_d$ is a fictitious disorder
potential, which contains a number of parameters. These parameters (Lagrange
multipliers) have to be tuned in such a way that some moments of the {\em a
posteriori} (annealed) distribution of the disorder are equal to the
{\em a priori} (quenched) ones. In annealed averages, the a posteriori
distribution $P^*(\{\la_i\},\{z_i\})$ is defined as

\beq
P^*(\{\la_i\},\{z_i\})\equiv{P(\la_i)\exp(-\be{\cal H}^*(\{\la_i\},\{z_i\}))
\over\dle{\cal Z}^*_N\dre}\ .
\label{P*}
\eeq

\noi The average over this distribution will be denoted by $\lgl\cdot\rgl
\equiv\int d\{\la_i\}\sum_{W_N}P^*(\{\la_i\},\{z_i\})\cdot\ \ $. In 
principle, one has to fix all the moments of $P^*_{\rm ann}(\{\la_i\})
\equiv\sum_{W_N}P^*(\{\la_i\},\{z_i\})$ to obtain the quenched result, 
which is as difficult as the direct computation of (\ref{fq}). 
Nevertheless, one can hope to obtain a reasonable approximation of the 
quenched case by fixing a few suitably chosen moments. Moreover, the method 
is variational, and fixing more and more moments yields thighter lower 
bounds for the quenched free energy.

In this work, we have considered three different cases of annealing: without
constraints ($a_0$), constraining the first moment of overall 
hydrophilicity ($a_1$), constraining the first and the second moment of 
overall hydrophilicity ($a_2$). For all these cases, we obtain the same 
formal expression for the effective homo-polymer partition function

\beq
{\cal Z}_N^{\rm eff}=\sum_{W_N}\exp\left[N\be_0+\be_1\sum_i z_i+\be_2\sum_i
z_i^2\right ]\ ,
\label{b3}
\eeq

\noi and any further complexity is hidden in the computation of $\be_0$, 
$\be_1$ and $\be_2$ for the different cases.
The strategy we will follow, is to study the general homo-polymer model
defined by (\ref{b3}), in the $(\be_1,\be_2)$-plane (the $\beta_0$
dependence being trivial). Then, we investigate to which temperature 
dependent trajectories in the $(\be_1,\be_2)$-plane, the three annealed 
averages give rise.

In case ($a_0$), the simple annealed average (\ref{mZ}) has already been 
computed in the preceding section, and equation (\ref{b3}) is recovered 
with the definitions

\beq
\be_0=1            \ ,\quad\quad
\be_1=\be\la_0     \ ,\quad\quad
\be_2=\ha\be^2\la^2\ .
\label{b30}
\eeq

\noi
We can immediately argue from equation (\ref{b3}) and (\ref{b30}) that
even hydrophobic chains ($\la_0<0$) will be swollen at low enough temperature.
Indeed, since $\be_2\gg\left |\be_1\right |$ as $\be\to
\infty$, the
repulsive $\be_2\sum_iz_i^2$ term causes the number of contacts with solvent
to be maximized, independently of the sign and the value of $\la_0$.

In case ($a_1$), we fix the a posteriori overall hydrophilicity 
$\sum_i\la_i/N$ to its a priori value $\la_0$:

\beq
{\lgl\sum_i\la_i\rgl\over N}=\la_0\ .
\label{l0}
\eeq

\noi We impose this constraint by defining a generalized partition
function ${\cal Z}_N^{a_1}(\{\la_i\},\hh)$ which depends on the Lagrange
multiplier $\hh$, and by finding the effective value $\hh^*$ for which
(\ref{l0}) holds:

\beq
{\cal Z}_N^{a_1}(\{\la_i\} ,\hh)=\sum_{W_N}\exp\left[\be\sum_i\la_i
z_i-\be \hh\left(\sum_i\la_i-N\la_0\right)\right]\ .
\label{Z1}
\eeq

\noi We recover (\ref{b3}) with the following definitions 

\beq
\be_0={\be^2\la^2\hh^2\over2}\ ,\quad\quad
\be_1=\be\la_0-\be^2\la^2\hh          \ ,\quad\quad
\be_2={\be^2\la^2\over2}\ .
\label{b31}
\eeq

\noi In terms of $\hh$, condition (\ref{l0}) becomes

\beq
\hh^*={\lgl\sum_iz_i\rgl(\hh^*)\over N}\ ,
\label{hc}
\eeq

\noi and the free energy $f_{a_1}$ is thus given by

\beq
f_{a_1}=-{1\over\be N}\ln\dll{\cal Z}_N^{a_1}(\hh^*)\drl\ .
\label{fa1}
\eeq

\noi Note that the quenched free energy computed in the continuum model by 
Garel {\it et al.} \cite{GLO}, is exactly the free energy $f_{a_1}$ 
(\ref{fa1}), if one performs the annealed average with constraint 
(\ref{l0}) within the analytic calculation scheme of reference \cite{GLO}. 
We will comment on this in the final discussion of section \ref{disc}.

In case ($a_2$), in addition to (\ref{l0}), we also put a constraint on the
overall variance $\sum_i(\la_i-\la_0)^2/N$:

\beq
{\lgl\sum_i\la_i^2\rgl\over N}=\la^2+\la_0^2\ .
\label{l2}
\eeq

\noi In the same way as before, we introduce a second Lagrange
multiplier $\hs$, and we define a generalized partition function

\beq
Z_N^{a_2}(\{\la_i\},\hh,\hs)=\sum_{W_N}\exp\left[\be\sum_i\la_iz_i-
\be\hh\left(\sum_i\la_i-N\la_0\right)-{\be\hs\over2}\left(\sum_i\la_i^2-
N(\la^2+\la_0^2)\right)\right]\ .
\label{Z2}
\eeq

\noi After performing the average, we recover (\ref{b3}) with the following 
definitions 

\beq
\be_0=\ha\left({\be^2\la^2\over s'}(\hh\!+\!\la_0\hs)^2\!+\!\be\la^2\hs
\!-\!\ln s'\right)\ ,\quad
\be_1={\be\la_0\!-\!\be^2\la^2\hh\over s'}\ ,\quad
\be_2={\be^2\la^2\over2s'}\ ;\quad\quad
s'=1\!+\!\be\la^2\hs \ .
\label{b32}
\eeq

\noi Conditions (\ref{l0}) and (\ref{l2}) yield two coupled equations for
$\hh^*$ and $\hs^*$ with solution:

\begin{eqnarray}
\hh^*&+&\la_0 \hs^*=\De_2(\hh^*,\hs^*)\ , \label{h2c}\\
\hs^*&=&{\sqrt{1+4\be^2\la^2(\De_3(\hh^*,\hs^*)-\De_2^2(\hh^*,\hs^*))}
-1\over2\be\la^2}
\label{s2c}\ ,
\end{eqnarray}

\noi where $\De_2\equiv\lgl\sum_iz_i\rgl/N$ and $\De_3\equiv\lgl\sum_i
z_i^2\rgl/N$. In the next section, we will show that $\De_2$ and $\De_3$ are
closely connected to effective  $2$-, respectively $3$-body interactions
between the monomers.
The free energy $f_{a_2}$ is now given by

\beq
f_{a_2}=-{1\over\be N}\ln\dll{\cal Z}_N^{a_2}(\hh^*,\hs^*)\drl\ .
\label{fa2}
\eeq

\noi As mentioned before, it can be shown \cite{Ku} that at any temperature 
$f_{a_0}\le f_{a_1}\le f_{a_2}\le...\le f_q$. Hence, fixing more and more 
moments, we get a better approximation for the quenched free energy.

\section{Phase Diagram}
\label{phdg}

We now discuss the phase diagram of the model defined in (\ref{b3}), in
the $(\be_1,\be_2)$-plane, with $\be_0$ constant:

\beq
{\cal Z}_N^{\rm eff}=C\sum_{W_N}\exp\left[\be_1\sum_i z_i+
\be_2\sum_i z_i^2\right ]\ .
\label{b2}
\eeq

\noi The number $z_i$ of nearest-neighbor contacts of monomer $i$ (not at
the chain ends) with
solvent molecules can be expressed in terms of the number $n_i$ of
nearest-neighbor monomers of monomer $i$ not along the chain using
$n_i=2(d\!-\!1)-z_i$, where we have exploited the incompressibility condition. 
In this way the quantities $\sum_iz_i$ and $\sum_iz_i^2$ can be related
to effective 2-body and 3-body interactions between the monomers. We define 
$N_2$ to be the number of nearest-neighbor 2-monomer contacts not along 
the chain, while $N_3$ is the number of {\em nearest} 3-monomer contacts 
not along the chain. On a hypercubic lattice in $I\!\!R^d$, we have a 
nearest 3-monomer contact when two monomers are both nearest neighbors (not 
along the chain) to the third monomer and we have that

\beq
N_2={\sum_i n_i\over2}\ ,\hspace{2cm}
N_3={(\sum_i n_i^2\!-\!\sum_i n_i)\over2}\ .
\label{N2}
\eeq

\noi In terms of $N_2$ and $N_3$, the reduced Hamiltonian can be rewritten

\beq
-\be{\cal H}^{\rm eff}=2(d\!-\!1)[\be_1+2(d\!-\!1)\be_2]N-
[2\be_1+2(4d\!-\!5)\be_2]N_2+2\be_2 N_3\ .
\label{H2}
\eeq

\noi Since $\be_2\geq0$ for the models that we have considered, the 3-body 
term is either attractive or absent. Note that $\be_2$ enters also in the
2-body term with a {\em repulsive} effect; as already noted, the total
contribution of the $\be_2$ term in equation (\ref{b2}) has to be repulsive.
We recall that the self-avoidance constraint automatically introduces 
effective $n$-body repulsive terms, with $n=2,3,4$ and so on.

For $\be_2=0$, we have a self-avoiding walk with only $2$-body
interactions, each of energy $2\be_1/\be$. For these models, the existence of 
a critical value $\be_1^*<0$ at which the chain undergoes a second order 
$\theta$-transition, is well known \cite{DeGen1,Sal,SenSte}. The transition 
takes place when the $2$-body attractive interaction exactly balances the 
$2$-body steric repulsion. For $\be_1>\be_1^*$, the chain is in the swollen 
phase, while for $\be_1<\be_1^*$, the chain is in the collapsed phase.
The $\theta$-point is a tri-critical point \cite{DeGen2} corresponding to
a $\phi^6$ field theory with a Landau-Ginzburg functional; the necessary
stabilizing $3$-body term is provided by the self-avoidance constraint.

When $\be_2>0$, the attractive $3$-body term $2\be_2 N_3$ competes with the
corresponding steric repulsion. At the mean field level \cite{ItzDro}, with
increasing 
$\be_2$, a tri-critical line departs from the $\theta$-point at $\be_2=0$. 
The tri-critical line ends at a multi-(tetra-)critical point $(\be_1^m,
\be_2^m)$, when the $3$-body attractive interaction exactly balances the 
steric repulsion. This corresponds to a $\phi^8$ Landau-Ginzburg theory, since
the necessary stabilizing $4$-body term is provided by the
self-avoidance constraint.
Increasing $\be_2$ further, the transition line becomes a coexistence line
between the swollen and the compact phase. This phase diagram is 
qualitatively sketched in Figure 1.

At zero temperature, when the entropy is negligible with respect to the
energy, we can give rigorous results for the asymptotic behavior of the
coexistence line. If $a=\be_1/\be_2$ is fixed and $\be_2\to\infty$,
we can rewrite the Hamiltonian

\beq
-\be{\cal H}^{\rm eff}=\be_2\sum_i(a z_i+z_i^2)=-N\be_2{a\over4}+\be_2
\sum_i\left({a\over2}+z_i\right)^2\ .
\label{Ha}
\eeq

\noi 
Since $0\!<\!z_i\!<\!2(d\!-\!1)$, for $a\!>\!-2(d\!-\!1)$, the ground 
state is at $z_i=2(d\!-\!1)$ for any $i$, and the walk is swollen.
\nl
On the other hand, for $a\!<\!-2(d\!-\!1)$, the ground state is at
$z_i=0$ for any $i$, and the walk is maximally compact.
\nl
For $a\!=\!-2(d\!-\!1)$ (i.e. $\be_1\sim-2(d\!-\!1)\be_2$), the energy
of the two competing ground states is the same, and there is phase
coexistence.

The presence of a multi-critical point, if not rigorously proved, is
numerically established, as one can see from Figure 2, where the order
parameter $\De_2(\be_1,\be_2)$ is plotted as a function of $\be_1$ for
different values of $\be_2$.

Camacho and Schanke \cite{CS}, using exact enumerations, have obtained
a phase diagram which exhibits similar features as our. 
A transition line between the swollen and the collapsed phase is present, 
which is first order at low temperatures, and becomes second order at higher 
temperatures through a multi-critical point.
However, they treat the quenched case and describe a slightly different
model (i.e. the HP-model). A translation of this model in terms of hydrophobic
charges would introduce an extra interaction term which depends on the
product of the charges, and is absent in our model.

\section{Transfer Matrix}
\label{tm}

We have addressed the numerical study of the lattice model defined by
(\ref{b3}), by the transfer matrix technique on a two-dimensional square
lattice. With this method it is possible to consider infinite polymers on
a lattice of finite width (strip) \cite{Kl,Der,SalDer}. The price to pay
is the uncertain extrapolation of the thermodynamic limit, caused by the
limited width of the strip that we can achieve.

In a grand-canonical context, the generalized two point correlation
function is defined as

\beq
{\cal G}(x,r,\be_1,\be_2)=\sum_{N=1}^{\infty}\sum_{W_N}x^N
\exp\left[\be_1\sum_i z_i+\be_2\sum_i z_i^2\right]\ ,
\label{G}
\eeq

\noi where $x$ is the step fugacity and the second sum runs over the
SAW of $N$ steps which connect the origin with an arbitrary point at
distance $r$. We have neglected the dependence on $\be_0$ because it only
affects a simple rescaling of fugacity. The two-point correlation
function decreases exponentially in $r$ at long distances, if $x$ is less
than the critical fugacity $x_c(\be_1,\be_2)$. This defines the correlation
length $\xi(x,\be_1,\be_2)$:

\beq
{\cal G}(x,r,\be_1,\be_2)\sim\exp\left(-{r\over\xi(x,\be_1,\be_2)}
\right)\ .
\label{xi}
\eeq

\noi
The correlation length $\xi_n(x,\be_1,\be_2)$ can be calculated exactly on 
a lattice strip of infinite length and finite width $n$, with the TM method.
The idea is to write recursion relations between a strip of length $r$ and 
a strip of length $r\!+\!1$. We consider a walk on the strip, which goes 
from the left to the right, and we cut the strip at column $r$. The local 
configuration at $r$ is then given by the set of occupied sites of column 
$r$ and how these are connected to each other by the part of the walk at 
the left of $r$. Since the interaction $\be_2\sum_i z_i^2$ gives rise to 
effective 3-body interactions, it is necessary to define the local 
configurations at stage $r$ taking the lattice bonds occupied by the walk 
between the columns $r\!-\!2$ and $r$ into account.
\nl
We combine all the possible local configurations $i$ at column $r$, with
all the possible local configurations $j$ at column $r\!+\!1$. They yield 
a non-zero TM element $T_{ij}$ if it is possible to connect them, without 
producing disconnected pieces, and $T_{ij}$ is given by

\beq
T_{ij}=x^{t_{ij}}\exp\left[\be_1\sum_{\alpha=1}^n z_{\alpha }^{ij}+\be_2
\sum_{\alpha=1}^n(z_{\alpha}^{ij})^2\right]\ ;
\label{Tij}
\eeq

\noi where $t_{ij}$ is the number of occupied bonds between columns 
$r\!-\!1$ and $r$, and $z_{\alpha}^{ij}$ is the number of non occupied 
nearest-neighbor sites of the site at row $\alpha$ and column $r\!-\!1$, if
this is occupied by the walk (see Figure $3$).

The number of possible configurations, and therefore the computational
effort, can be strongly reduced by considering periodic boundary
conditions (the strip becomes a cylinder) and then by exploiting all the
symmetry properties of the strip. Furthermore, periodic boundary conditions
reduce the finite size effects. In this way, within reasonable time, we are 
able to study strip widths up to $n=6$, corresponding to $5387$ 
configurations and $154149$ non-zero matrix elements.

The correlation function can be expressed in terms of the trace of the
$r$-th power of the TM $T$:

\beq
{\cal G}(x,r,\be_1,\be_2)\sim{\rm Tr}\ T^r\ ,
\label{Tr}
\eeq

\noi and the correlation length (\ref{xi}) is related to the largest 
eigenvalue $\la_n^{\rm max}(x,\be_1,\be_2)$ of $T$, for a strip of width 
$n$:

\beq
\xi_n(x,\be_1,\be_2)=-{1\over\ln\la_n^{\rm max}(x,\be_1,\be_2)}\ .
\label{xin}
\eeq

\noi The critical fugacity $x_c^n$ is determined by the value at which the
correlation length diverges, i.e.

\beq
\la_n^{max}(x_c^n,\be_1,\be_2)=1\ .
\label{lm}
\eeq

\noi The computation of the free energy per monomer
$f=-\ln{\cal Z}_N^{\rm eff}/ (\be N)$
and of any other quantity of physical interest (e.g. the mean
number of monomer-solvent contacts $\De_2$), is now straightforward in
terms of the critical fugacity:

\beq
f={1\over\be }\ln x_c\ ,\quad\quad
\De_2 =-{\partial\ln x_c\over\partial\be_1}\ ,\quad\quad
\De_3 =-{\partial\ln x_c\over\partial\be_2}\ .
\label{fdd}
\eeq

\noi
The thermal exponent $\nu$, which characterizes the divergence of the
correlation length at the critical fugacity: $\xi\sim(x_c-x)^{-\nu}$,
is a good indicator of a collapse transition. A SAW in two dimensions
has the value $\nu=3/4$ in the swollen phase \cite{Nien}, $\nu=1/2$ in the
collapsed phase and $\nu=4/7$ on the tri-critical line \cite{ConJMS,DupSal}.

In order to compute the thermal exponent, we first calculate the density 
$\rho_n(\be_1,\be_2)$ of monomers in the strip

\beq
\rho_n(\be_1,\be_2)=-{x_c^n (\be_1,\be_2)\over n}
{\partial\xi_n^{-1}(x_c^n,\be_1,\be_2)\over \partial x}\ .
\label{rn}
\eeq

\noi Then, we use a phenomenological renormalization (PR)
group procedure \cite{Night,DerSez} to obtain finite size estimates for 
the thermal exponent; the underlying hypothesis is the finite size scaling
behavior \cite{FisBar} of the correlation length for $n\!\gg\!1$ and
$\left(x_c-x\right)\!\ll\!1$:

\beq
\xi_n\left(x,\be_1,\be_2\right)
=n\; g\left[n^{1/\nu}\left(x_c-x\right),\be_1,\be_2\right]\ ,
\label{fsc}
\eeq

\noi where $g$ is a scaling function. Using the single strip critical
fugacity estimate (\ref{lm}) leads to

\beq
\nu_{n,n-1}=\left({\ln(\rho_n/\rho_{n-1})\over\ln(n/(n-1))}+2\right)^{-1}\ .
\label{nu}
\eeq

\noi Note that we compare the derivative of the correlation length at 
criticality for two consecutive strip widths, but criticality is determined
in a different way for different widths.

This is not the most accurate way of applying the ideas of the PR. In fact, 
the critical fugacity can be determined for two consecutive widths at once 
by

\beq
{\xi_n\left(x_c^{n,n-1},\be_1,\be_2\right)\over n}=
{\xi_{n-1}\left(x_c^{n,n-1},\be_1,\be_2\right)\over n-1}\ .
\label{xpr}
\eeq

\noi This estimate is better than (\ref{lm}) and the thermal exponent can
easily be obtained

\beq
\nu_{n,n-1}=\left({\ln({\partial\xi_n(x_c^{n,n-1},\be_1,\be_2)\over
\partial x}/{\partial\xi_{n-1}(x_c^{n,n-1},\be_1,\be_2)\over\partial x})
\over\ln(n/(n-1))}-1\right)^{-1}\ .
\label{nupr}
\eeq

\noi Nevertheless, we have used the rougher formulae (\ref{lm}) and 
(\ref{nu}), because solving (\ref{xpr}) numerically, is a much harder
task, especially in proximity of coexistence.

\section{Results}
\label{res}

We will show that the trajectory that the system follows in
the $(\be_1,\be_2)$-plane, for the different models with decreasing
temperature, only depends on the fraction $\la_0/\la\!\equiv\!\lae$. The
position on the trajectory at a given temperature, however, does depend on
$\la_0$ and $\la$ separately. Although the trajectories can not be
calculated analytically, we give a general qualitative picture, which will
be confirmed by the numerical data.

The maximum strip width avalaible, $n=6$, is rather small.
Nevertheless, the data in the various plots already show a nice convergence,
and we think that the obtained results are reliable.
Moreover, we are mainly interested in qualitative features of the phase 
diagram and not in precise quantitative values of critical exponents or 
transition temperatures.
   
\subsection{ Simple Annealing ($a_0$)}
\label{ann}
   
After elimination of the temperature in (\ref{b30}), the locus of
the trajectory in the $(\be_1,\be_2)$-plane is given by the equation

\beq
g_{a_0}(\be_1,\be_2)\equiv\be_2-{\la^2_{\rm eff}\be_1^2\over2}=0
\ ,
\label{ga0}
\eeq

\noi which describes a parabola. As the temperature is positive,
the $(\be_1\!>\!0)$-branch of this parabola has to be considered, for
$\lae\!>\!0$, while the $(\be_1\!<\!0)$-branch is relevant for
$\lae\!<\!0$.

\noi Hence, at sufficiently low temperature, the chain will be always
swollen, no matter how strongly hydrophobic $\lae$ is (i.e. $\lae<0$).
We note here that
the opposite result, i.e. even highly hydrophilic chains are compact at
sufficiently low temperature, has been found by Garel {\it et al.} \cite{GLO}
in the corresponding continuum model, due to improper consideration of the
incompressibility condition. This condition is automatically accounted for
in the definition of our lattice model.

The typical behaviour for a strongly hydrophobic chain ($\lae\!\ll\!-1$) is
as follows. At high temperatures, it will be swollen for entropic reasons.
Then with decreasing temperature, it will undergo a 2nd order
$\theta$-transition from swollen to collapsed. Finally, at even lower
temperature, it will undergo a 1st order collapsed to swollen transition.
We present a numerical evidence of this remarkable re-entrant behaviour in
Figure 4: the crossings of the various $n$-estimates of the thermal
exponent are typical of $\theta$-point \cite{Sal} and are just
around the value $\nu_{\theta}\simeq0.57$, and the jump of the order
parameter $\De_2=\lgl\sum_iz_i/N\rgl$ provides strong evidence for the
first order transition from the compact to the swollen phase; the
value of the compact phase being $\De_2\simeq0$, and in the swollen
phase $\De_2\simeq2$.
We have thus shown that considering $3$-body interaction does not change
the universality class of the $\theta$-transition, as long as one is referring
to the tri-critical line. Although not surprising, this result is not
trivial in two dimensions. A more interesting theoretical question concerns
the value of the thermal exponent in $d=2$ at the multi-critical point, but
the TM approach employed here, is uneffective because of the
limited strip width we are able to study.

We note here that we have considered the overall hydrophilicity $\la_0$ to
be constant. The chemical reactions that give rise to annealed 
hydrophilicities, however, may be temperature dependent and may cause 
$\la_0$ to vary with temperature \cite{pr}. Nevertheless, the re-entrant 
behavior is a quite robust feature: even when $\lambda_0$ diverges 
exponentially to $-\infty$ ($\la_0^*\!<\!0$) with a rate $\alpha\!>\!0$

\beq
\la_0\left(\be\right)=\la_0^*\exp\left(\alpha\be\right)\ ,
\label{l0v}
\eeq

\noi re-entrant behavior is still observed for $\alpha$ not too big.

\subsection{ Fixing the Mean ($a_1$)}
\label{fixm}

After elimination of the temperature in (\ref{b31}), the locus of the
trajectory in the $(\be_1,\be_2)$-plane is given by the equation

\beq
g_{a_1}(\be_1,\be_2)\equiv\be_1-\sqrt{2\be_2}\lae+
2\be_2\De_2(\be_1,\be_2)=0\ ,
\label{ga1}
\eeq

\noi which has to be combined with condition (\ref{hc}) (fixing the
mean)

\beq
\De_2(\be)\equiv
\De_2(\be_1,\be_2)={1\over N}{\partial\ln{\cal Z}_N^{\rm eff}
(\be_1,\be_2)\over\partial\be_1}={1\over N}\lgl\sum_iz_i
\rgl(\be_1,\be_2)\ .
\label{D2}
\eeq

\noi
First, we show that equation (\ref{ga1}) defines a unique trajectory
$\be_1^{tr}(\be_2)$. It is easy to see that $\partial g_{a_1}/\partial
\be_1>0,\ \forall \be_1,\ \be_2$. Hence, we can apply the implicit function
theorem, but only if $\be_2<\be_2^m$, when $g_{a_1}(\be_1,\be_2)$ is a
continuous function of its arguments.
For $\be_2>\be_2^m$, $g_{a_1}(\be_1,\be_2)$ (in particular $\De_2$), is 
discontinuous on the coexistence line $\be_1^{co}(\be_2)$. Since the
discontinuity is developed in the thermodynamic limit, and since 
$\partial g_{a_1}/\partial\be_1>0$ for any finite $N$, the only possibility 
at coexistence is $g_{sw}\equiv g_{a_1}(\be_1\downarrow\be^{co}_1(\be_2),
\be_2)>g_c\equiv g_{a_1}(\be_1\uparrow \be^{co}_1(\be_2),\be_2)$ 
(the chain is collapsed for $\be_1<\be^{co}_1(\be_2)$ and swollen 
otherwise). If $g_{sw}$ and $g_c$ have the same sign, (\ref{ga1}) is still 
uniquely satisfied far away from coexistence. Instead, if $g_{sw}\!>\!0$ 
and $g_c\!<\!0$, (\ref{ga1}) can only be satisfied on the coexistence line.
In this case, a fraction $f_c$ of the chain is collapsed and the remaining 
fraction $1-f_c$ is swollen, so that physical mean values are mixtures of 
the pure phase values:

\beq
\De_2(\be)=f_c\ \De_{2,c}(\be)+(1\!-\!f_c)\ \De_{2,sw}(\be)\ .
\label{fcd}
\eeq

\noi Thus, equation (\ref{ga1}) becomes a condition on $f_c$:

\beq
f_c\ g_c+(1\!-\!f_c)\ g_{sw}=0\ .
\label{fcg}
\eeq

\noi
We now prove that at zero temperature ($\be_2\to\infty$) the only way to
satisfy (\ref{ga1}) for the chain, is to be at the coexistence line with
$f_c=\ha$, i.e. half collapsed and half swollen, independently of the
value of $\lae$, as far it is kept fixed. For $\be_1/a\simeq\be_2\to\infty$,
(\ref{ga1}) becomes

\beq
a+2\De_2(\infty)=0\ .
\label{a1}
\eeq

\noi As we have seen at the end of section \ref{phdg},
for $a\!>\!-2 $ the chain is swollen, and $\De_{2,sw}(\infty)=2$, but
(\ref{a1}) implies $a=-4$ which is a contradiction. Similarly for
$a\!<\!-2$, the chain is compact and $\De_{2,c}(\infty)=0$, implying
$a\!=\!0$. Hence, the only remaining possibility is $a\!=\!-2$, i.e.
coexistence of the swollen and the compact phase:

\beq
\De_2(\infty)=f_c\De_{2,c}+(1\!-\!f_c)\De_{2,sw}\ ,
\label{D2i}
\eeq

\noi where $f_c$ is the collapsed fraction of the chain at coexistence.
Plugging this in (\ref{a1}), yields $f_c=\ha$, such that $\De_2(\infty)=1$.

The phase separation already occurs at finite temperature, since condition 
(\ref{ga1}) implies that $\De_2(\be_1,\be_2)$ is a continuous function 
along the trajectory, and since the only way to reach the value 
$\De_2(\infty)=1$ continuously, is to move along the coexistence line.

All this, in combination with the numerical data, gives the following
qualitative scenario of what happens lowering the temperature:

$\bullet$ There exists a particular value $\la_m\!<\!0$ such that for
$\lae=\la_m$ the
trajectory passes through the multi-critical point, and then follows the
coexistence line.

$\bullet$ For $\lae>\la_m$, the trajectory hits the coexistence line
coming from the swollen phase, and this will happen further away from the
multi-critical point the larger $\lae-\la_m$ is. Then, it follows the
coexistence line, and the collapsed fraction ($f_c$) of the chain steadily
increases to become $\ha$ at zero temperature.

$\bullet$ For $\lae<\la_m$, the trajectory hits the coexistence line
coming from the compact phase.  This means that the chain first collapses
with a 2nd order $\theta$-transition, before the trajectory hits the
coexistence line. Then, it follows the coexistence line, and the collapsed
fraction ($f_c$) of the chain steadily decreases to become $\ha$ at zero
temperature.

This qualitative scenario is confirmed by the numerical evidence shown
in Figures $5$ and $6$. They respectively show the numerical results for 
the trajectory in the ($\be_1,\be_2$)-plane, and the variation of the 
order parameter $\De_2(\infty)$ with temperature. 

\subsection{ Fixing the Mean and the Variance ($a_2$)}
\label{fixmv}

Using (\ref{h2c}) and (\ref{b32}), we obtain $\la\be=(\be_1+2\De_2
\be_2)/\lae\equiv\chi>0$, and the locus of the trajectory in the 
($\be_1$,$\be_2$) plane is given by the following equation

\beq
g_{a_2}(\be_1,\be_2)\equiv{\chi^2\over2\be_2}-{1\over2}
\sqrt{{1\over4}+\chi^2\left(\De_3-\De_2^2\right)}=0\ ,
\label{ga2}
\eeq

\noi where $\De_2(\be)$ is defined as in (\ref{D2}), and

\beq
\De_3(\be)\equiv\De_3(\be_1,\be_2)=
{1\over N}{\partial\ln{\cal Z}_N^{\rm eff}(\be_1,\be_2)\over\partial\be_2}=
{1\over N}\lgl\sum_iz_i^2\rgl(\be_1,\be_2)\ .
\label{D3}
\eeq

\noi The qualitative behaviour of the trajectories in the phase plane is
very similar to that of the previous subsection. The collapsed fraction
$f_c$ of the chain, however, does depend on $\lae$ at zero temperature.
In order to show this, we repeat the same argument as before. For 
$\be_1/a\equiv\be_2\to\infty$, condition (\ref{ga2}) simplifies to

\beq
a+2\left(\De_2(\infty)-\lae\sqrt{\De_3(\infty)-\De^2_2(\infty)}\right)=0\ ,
\label{a2}
\eeq

\noi
which is the analogous of equation (\ref{a1}).
For $a\!>\!-2$ the chain is swollen and $\De_{2,sw}(\infty)=2$,
$\De_{3,sw}(\infty)=4$, but (\ref{a2}) implies $a=-4$ which is a
contradiction. For $a\!<\!-2$ the chain is collapsed, and
$\De_{2,c}(\infty) =\De_{3,c}(\infty)=0$ leads to $a=0$.
Again, we conclude that the chain is at coexistence at zero temperature,
but in this case using (\ref{D2i}) and the analogous formula for
$\De_3(\infty)$, we get the following condition for the collapsed
fraction $f_c$ of the chain:

\beq
f_c=\ha\ \left[1-{\lae\over\sqrt{1+\la^2_{\rm eff}}}\right]\ .
\label{fc2}
\eeq

\noi
In Figure $7$ the numerical results are shown for the variation of the order
parameter $\De_2$ with temperature. The analogous of Figure $6$ with the
trajectories in the ($\be_1,\be_2$)-plane turns out to be indistinguishable 
from Figure $6$ itself, and is therefore not shown.

The behaviour of the chain seems qualitatively unchanged adding the
constraint on the variance with respect to the fixed mean case. It can
easily be verified that for $\la_0=0$ equations (\ref{ga1}) and
(\ref{ga2}), defining the trajectories in the ($\be_1,\be_2$)-plane, are
equal, and they are qualitatively very similar for $\la_0\neq0$.
If we compare the free energies (\ref{fa1}) and (\ref{fa2}), however,
taking the proper values of $\be_0$ into account, we find that the
constraint on the variance is crucial for the low temperature behaviour of
the free energy. In the fixed mean case (like in the simple annealed case),
the free energy diverges linearly to $-\infty$ with $\be$, whereas fixing
also the variance yields a finite free energy. The divergence can easily be
understood. A fraction (all, in the simple annealed case) of the monomers
want to be as hydrophilic as possible and to maximize their solvent
contacts, in order to minimize the energy, while the other fraction has to
be very hydrophobic to keep the mean finite.
For entropical reasons the fractions are exactly $\ha$.
This is illustrated in Figure $8$, where the free energies are compared
for the various cases for the same values of $\la_0$ and $\la$.

\subsection{Towards the Quenched Average}

So far, we have only fixed overall moments of the type $\sum_i\la^l_i,
\ l\in I\!\!N$. In this way, we do not impose the $\la_i$ to be independent
variables. Or equivalently, even if we fix all the overall moments, the
$\la_i$ still have the complete freedom to rearrange themselves along the
chain. Hence, we can assume that fixing both mean and variance may be a
good approximation for a hetero-polymer, whose hydrophobicities are fixed,
but are allowed to migrate. In a protein, however, not only the
hydrophilicities, but also the positions along the chain are fixed. This
corresponds to the quenched case.

In order to get a reasonable approximation by means of annealed averages,
one should also ensure the independence of the $\la_i$. A first try might
be to impose e.g. $\lgl\sum_i\la_i\la_{i+1}\rgl/N=\la_0^2$, but after
performing the average, one discovers immediately that this would involve
a coupling between all the $z_i$, which, obviously, can not be done by the
TM method. Instead as a first approach, one could start with

\begin{eqnarray}
\dll\exp(-\ha\sum_{j,k}(\la_j\!-\!\la_0)M_{jk}(\la_k\!-\!\la_0)\!+\!\sum_j
\la_j\be(z_j\!-\!\hh)-\log(P(\{\la_i\}))\drl\ ,\nonumber \\
M_{jk}\equiv
\int_0^{2\pi}{dq\over2\pi}{\exp(iq(j\!-\!k))\over(\hs\!+\!\hd\cos(q))}.
\label{c1}
\end{eqnarray}

\noi After performing the average, this results in the following
expression (up to constants)

\begin{eqnarray}
\exp\!\!\left({\be^2\over2}\sum_{j,k}(z_j\!-\!\hh)M^{-1}_{jk}(z_k\!-\!\hh)\!
+\!\la_0\be\sum_j(z_j\!-\!\hh)\!+\!{N\over2}\log\left[\hs\!+\!\sqrt{\hs^2\!
-\!\hd^2}\ \right]\!\right)\!,\nonumber \\
M^{-1}_{jk}\equiv\hs\ \delta_{j,k}\!+
\!{\hd\over2}(\delta_{j,k-1}\!+\!\delta_{j,k+1})\ .
\label{c2}
\end{eqnarray}

\noi We have introduced the Lagrange multipliers $\hh$, $\hs$ and $\hd$,
which combined fix $\lgl\sum_j\la_j\rgl=N\la_0$, $\lgl\sum_j\la_j^2\rgl
=N(\la_0^2+\la^2)$, and $\lgl\sum_{j>k}\la_jM_{jk}\la_k\rgl=
\la^2_0\sum_{j>k}M_{jk}$, which ensures the independence of a 
linear combination of the $\la_j$. We expect that fixing the latter,
may already qualitatively describe the quenched case very well. The only
reason we did not do the numerics of this case, is of a purely practical
nature. In order to calculate $\lgl\sum_iz_iz_{i+1}\rgl$, we have to
consider configurations on 3 colums instead of on 2. This increases the
size of the transfer matrix so drastically that we would have to limit
ourselves to very narrow strip widths. Furthermore, we have an extra self
consistency equation (i.e. for $\hd$) to solve numerically. All this makes
it unfeasible (in terms of CPU time) for us at the moment, to perform this
calculation for a reasonable strip width (i.e. $\geq4$).

Nevertheless, one may anticipate that some of the typical behaviour found
for annealed averages, should not be present for the quenched case.
The re-entrant
behaviour at intermediate temperatures is due to the competition between
the configurational entropy and the energy on the one hand, and the entropy
of the $\la_i$ distribution on the other hand. In the quenched case, the
entropy of the $\la_i$ distribution is absent, and hence re-entrant
behaviour, if present at all, can not have its origin there.
The phase separation (in low dimensions) is due to the possibility for the
monomers to rearrange and to form a hydrophobic compact core. Since this
is not possible for the quenched polymer, we do not expect macroscopic phase
separation in that case.
Instead, microscopic phase separation seems to play an important role for
quenched sequences \cite{Tim,Mos}.
However, one might expect the quenched polymer to behave as an effective
homo-polymer, where the groundstate is either swollen or compact, depending
on the value $\lae$.

\section{Discussion of results}
\label{disc}

We have studied a simple lattice model for a random hydrophobic-hydrophilic
chain in a solvent, with a Gaussian distribution for the hydrophilicities.
We have considered the case of annealed disorder, without constraints, and
with constraints on the first and second moments of the overall hydrophobicity.

We have obtained both exact analytical results (mainly at $T=0$), and
numerical ones, employing the transfer matrix technique on a $2d$ square
lattice. One may ask whether the $2d$ results are relevant in $3d$ too.
For example, in the random sequence model with charge product
interaction, a simple mean field argument shows that $d=2$
is a very peculiar case \cite{SFA}.
For the considered model, analytical results at the mean field level 
\cite{GLO} do not show any qualitative difference between different spatial 
dimensions, and our exact results at $T=0$ exhibit the same 
qualitative behavior for any $d>1$. Hence, we believe the TM results in 
$2d$ to be meaningful in any dimension $d>1$.

We now discuss our results and compare them with the ones obtained by
Garel {\it et al.} \cite{GLO} in the corresponding continuum model.
The main result of \cite{GLO} is the fact that the annealed and quenched 
cases are very similar. They find that, at sufficiently low temperature, 
the polymer is always collapsed, even for hydrophilic chains $\la_0>0$. 
Depending on the average degree of hydrophobicity, the transition to the 
collapsed phase is either first or second order.
\nl

$\bullet$ For the simple annealed case ($a_0$), we have shown that for any 
hydrophobicity $\la_0>0$ the chain is always swollen, while for any 
$\la_0<0$, the chain is swollen at sufficiently low temperature. 
Using transfer matrix techniques, we have found that, for sufficiently 
negative $\la_0$, a temperature interval $(T_1,T_2)$ exists, where the 
chain is collapsed. Coming from the high temperature region the chain 
undergoes a standard 2nd order $\theta$-point transition at $T_2$ (in the 
same universality class as homo-polymers \cite{DupSal}). Lowering the 
temperature further, the chain undergoes a 1st order (re-entrant) 
transition at $T_1$ towards the swollen phase.

Hence, in the annealed case, we come to the opposite conclusion for the low 
temperature behavior as predicted by Garel {\it et al.} \cite{GLO}. 
Nevertheless, if one takes the incompressibility of the monomer-solvent 
system properly into account (i.e. putting an upper bound on the monomer 
density), it is possible to recover the same qualitative picture in the 
continuum model of \cite{GLO}, too.
\nl

$\bullet$
In the annealed case with fixing the mean ($a_1$), we have found that, for 
any $\lae$, there is coexistence of the swollen and the collapsed phase
(phase separation) at 
sufficiently low temperature, and the chain is exactly half collapsed and 
half swollen at $T=0$. 
For $\lae<\la_m$, a temperature interval $(T_1,T_2)$ exists where the chain 
is collapsed. At $T_2$ the chain undergoes a 2nd order $\theta$-transition 
from the swollen phase, while at $T_1$ a fraction of the chain swells, and 
lowering the temperature, the swollen fraction steadily increases to the 
$T=0$ value $\ha$. For $\lae>\la_m$ a temperature $T_1$ exists above which 
the chain is swollen. At $T_1$ a fraction of the chain collapses, and 
lowering the temperature, the collapsed fraction of the chain increases 
steadily to the $T=0$ value $\ha$.

As already noted in section \ref{ConAnn}, the expressions for the quenched 
case of \cite{GLO} are exactly the same as the ones we obtain for the
continuum model in the case ($a_1$). Using a one-parameter Gaussian trial 
wave function for the monomer density, Garel {\it et al.} \cite{GLO} find a
collapsed phase for any $\lae$ at low temperature. Inspired by the 
observation that they in fact describe the case ($a_1$), and by the phase 
separation observed in our lattice model, we tried a one-parameter trial 
function with a hydrophobic compact core (fraction $f_c$) and hydrophilic 
swollen tails (phase separation). In the low temperature limit, we recover 
the result $f_c=\ha$, and obtain a free energy that is considerably lower
than the one obtained using the trial function of \cite{GLO}. 

Note that the free energy diverges linearly to $-\infty$ for $T\rightarrow0$,
as in the annealed case (see Figure $8$). The same kind of divergence
appears in the quenched free energy computed in \cite{GLO}, but the quenched
free energy should not diverge at zero temperature.
\nl

$\bullet$
In the annealed case with fixed mean and variance ($a_2$), we have found
very similar results as for the case ($a_1$). The main differences are that 
the collapsed fraction of the chain at $T=0$ depends on $\lae$ (\ref{fc2}), 
and that the free energy does not diverge at $T=0$.
We repeated these calculations for the continuum model, and again we find 
that at $T=0$ the phase separation trial function yields a finite 
groundstate energy lower than the one obtained with a Gaussian trial function,
and the collapsed fraction of the chain ($f_c$) is 
found to be exactly (\ref{fc2}).
\nl

We conclude that, on the one hand we have good evidence that lattice and 
continuum models exhibit the same qualitative behavior, if constrained 
annealing is considered.
On the other hand, a good equilibrium description in the case of quenched
disorder for this model seems to be lacking at present, and carrying on 
the constrained annealing approximation procedure (e.g. by fixing 
correlations between different hydrophilicities, as explained in the
previous section) may be one way to address the problem of quenched 
disorder.

\section*{Acknowledgements}

We would like to thank H.~Orland for stimulating discussions.





\vskip 1truecm

\centerline{\bf Figure Captions}

\vskip 1truecm

{\bf Fig. 1}. Qualitative phase diagram in the ($\be_1$,$\be_2$) plane:
the solid line is the tri-critical $\theta$-line which ends in the
multi-critical point; the dashed line is the coexistence line.

\vskip 0.5truecm

{\bf Fig. 2}. Mean number of monomer-solvent close contacts
$\De_2\left(\be_1,\be_2\right)$ at varying $\be_1$ for different fixed values
of $\be_2$, with strip width from $2$ to $6$.
The compact-to-swollen transition is continuous for $\be_2<\be_2^m$ and
first order for $\beta_2>\be_2^m$, with the (very raw) estimate
$\be_2^m\simeq 0.75$.

\vskip 0.5truecm

{\bf Fig. 3}. Example of a transfer matrix element. Empty circles are solvent
molecules and dashed lines show the nearest-neighbor monomer-solvent
contacts. Configuration $i$ is defined at column $r$ and takes into account
how the walk steps back to column $r-2$ (solid line in ($b$)).
Configuration $j$ is defined at column $r+1$ and takes into account how the
walk steps back to column $r-1$, and thus, partially overlaps with
configuration $i$. The dotted line in ($b$) shows the non overlapping part of
configuration $j$. In this example the quantities needed for the computation
of the matrix elements (\ref{Tij}) are $t_{ij}=5$ and
$z^{ij}=(0,0,2,0,2,0)$ (the sites of column $r-1$ are ordered from the top
to the bottom of the strip).

\vskip 0.5truecm

{\bf Fig. 4}. Mean number of monomer-solvent close contacts $\De_{2,n}$
and thermal exponent $\nu_{n,n-1}$ at varying temperature, with strip
width $n$ from $2$ to $6$, in the case of unconstrained annealing with
$\la_0=-1$ and $\la=0.7$.
Evidence is provided for a second order swollen-to-compact $\theta$-transition
(see the crossings of different $n$-estimates of the thermal exponent around
the $\theta$-value $\nu_{\theta}\simeq0.57$), and for a first order
compact-to-swollen transition (see the abrupt jump of the order parameter
$\De_{2,n}$). The thermal exponent strongly fluctuates at the first order
transition due to the phenomenological renormalization method employed for its
calculation.

\vskip 0.5truecm

{\bf Fig. 5}. Mean number of monomer-solvent close contacts $\De_{2,n}$
at varying temperature, with strip width $n$ from $2$ to $5$, in the fixed mean
case. The behavior of the order parameter is the same (swollen at high
temperature and then $\theta$-transition to the collapsed phase) as in the
annealed case (see Figure $3$) until coexistence starts at $\be\simeq1.7$
and $\De_{2,n}$ starts increasing slowly; the asymptotic
($\be\rightarrow\infty$) value is $\De_2=1$.
The behavior of the thermal exponent is the same as in the
annealed case until coexistence starts. At coexistence the thermal exponent
should be the same as in the swollen phase, as far as a finite fraction of
the chain is swollen, but due to limited numerical precision we get higly
fluctuating values.

\vskip 0.5truecm

{\bf Fig. 6}. Trajectories in the ($\be_1$,$\be_2$) plane for different
values of $\lae$ in the fixed mean case, with strip width $n=5$.
The transition line has been located at the crossing of
two consecutive $n$-estimates of $\De_2$ (see Figure 2).

\vskip 0.5truecm

{\bf Fig. 7}. As in Figure $5$, in the fixed mean and variance case.
The asymptotic ($\be\rightarrow\infty$) value is $\De_2\simeq0.18$ (see
equation (\ref{fc2})).

\vskip 0.5truecm

{\bf Fig. 8}. Free energies $f_{a_0}$, $f_{a_1}$ and $f_{a_2}$, of the three
considered annealed cases, at varying temperature, with the same $\lae$, and
strip width $n=5$. While $f_{a_0}$ and $f_{a_1}$ diverge linearly to $-\infty$
as $\be\rightarrow\infty$, $f_{a_2}$ is constant in the same limit.

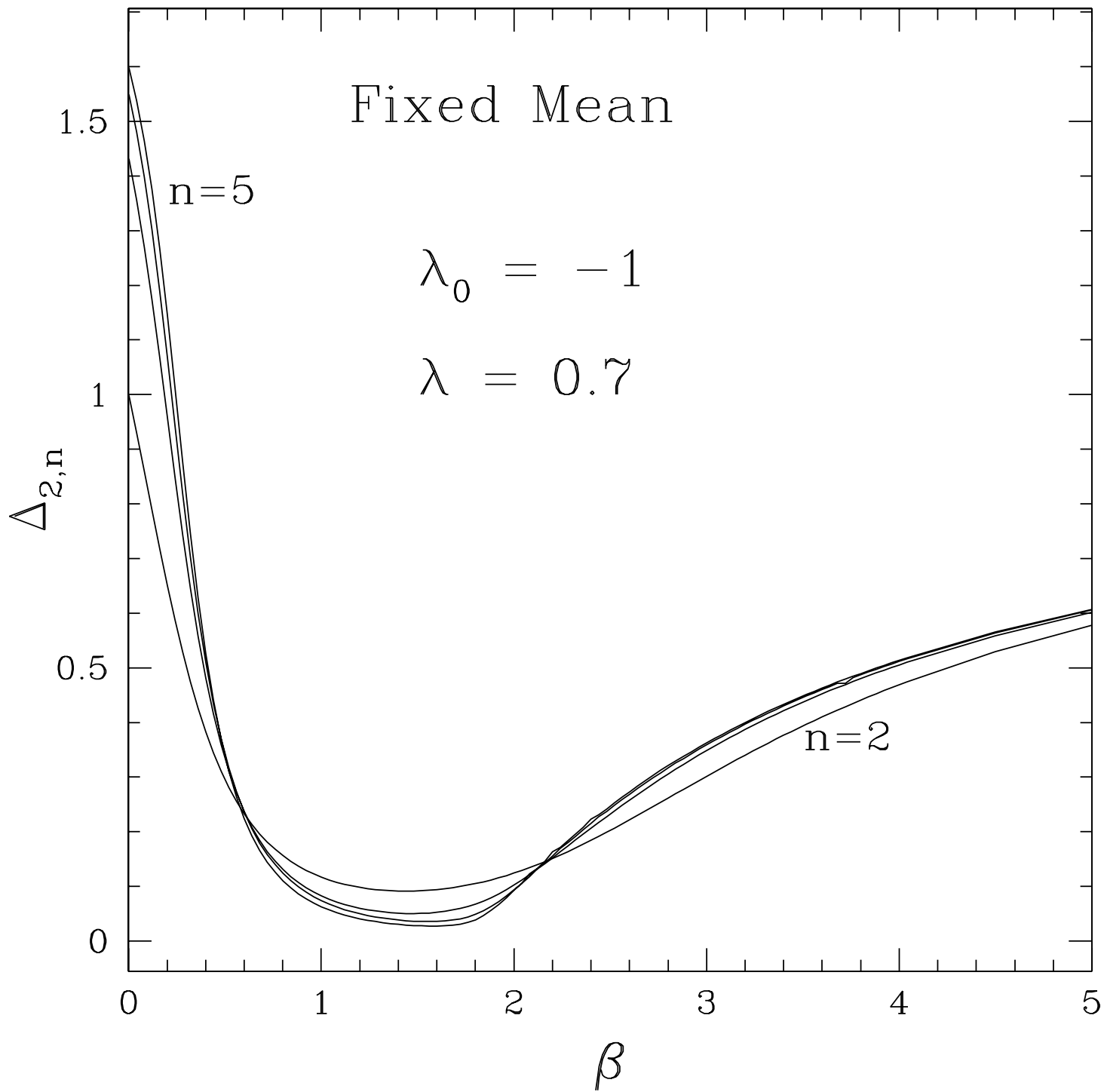
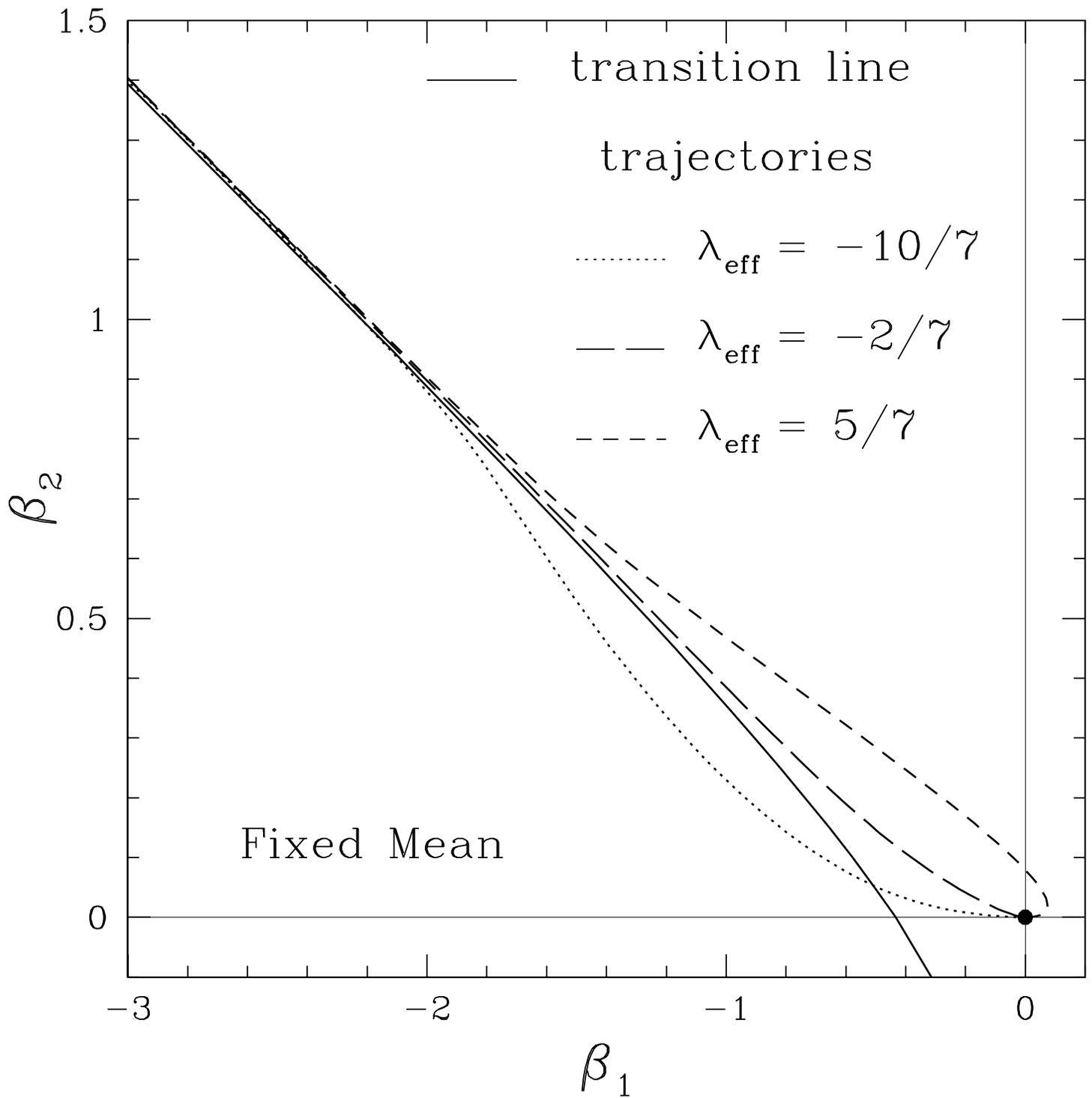
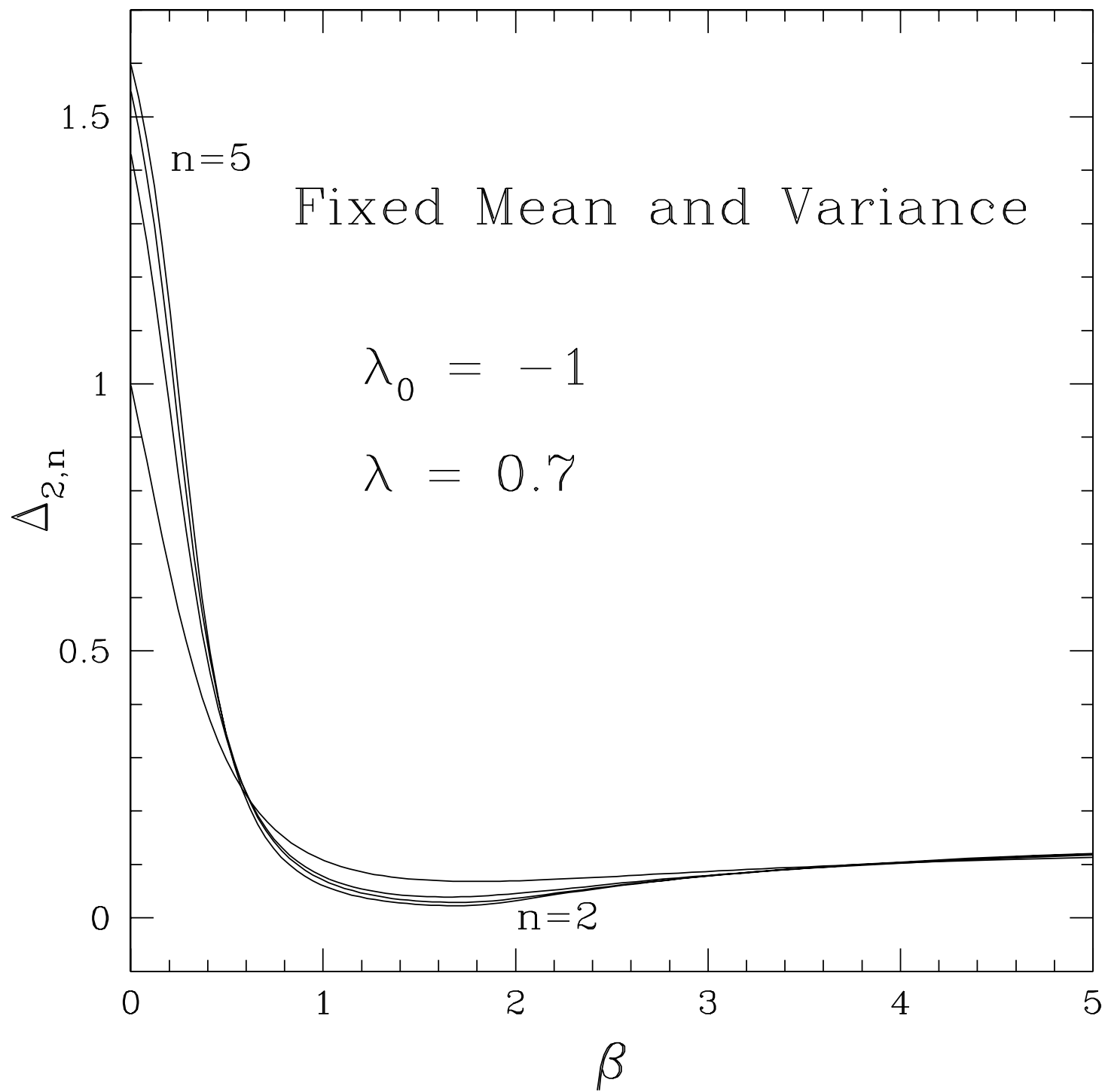
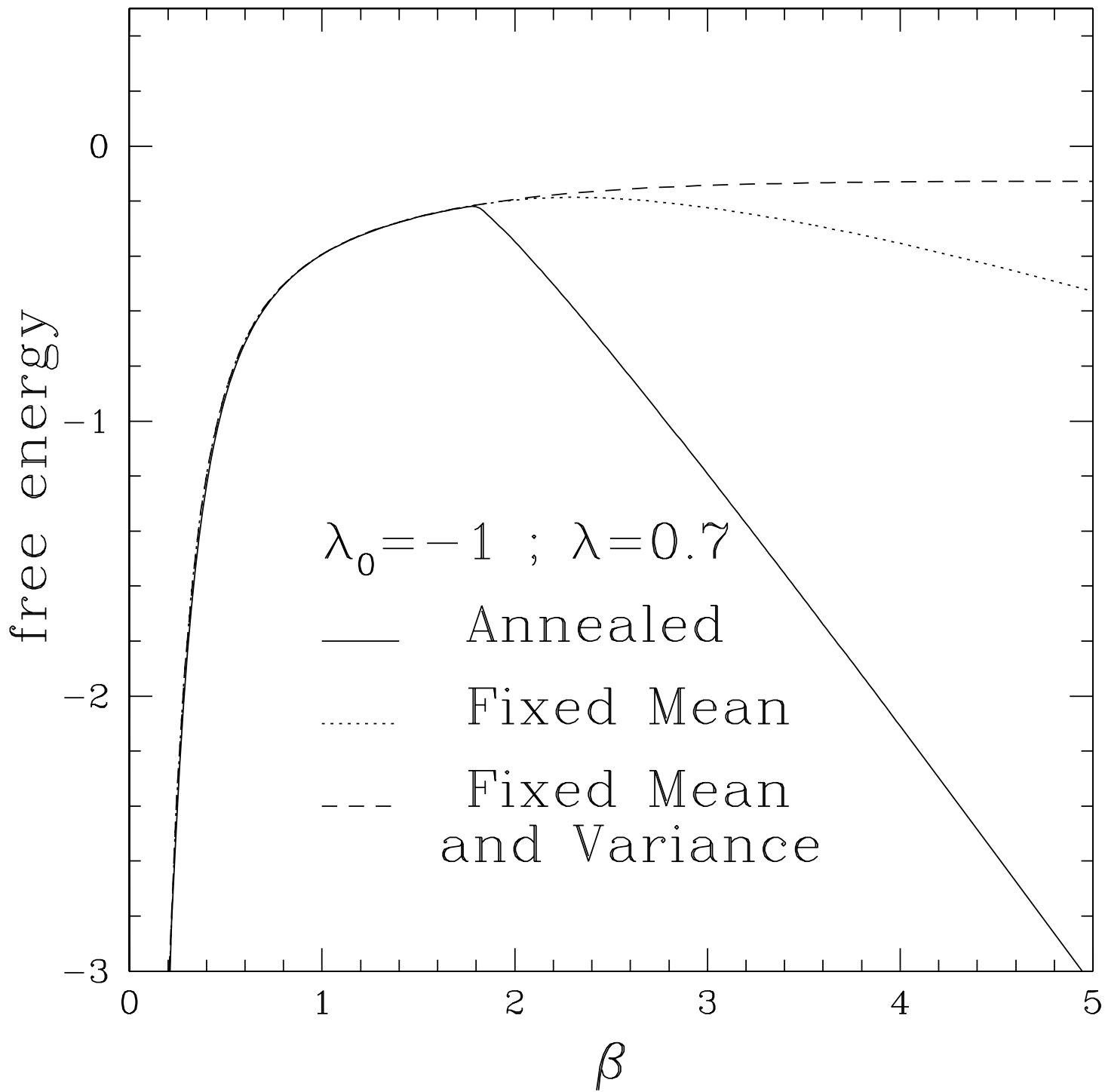
\end{document}